\documentclass{WileyMSP-template}
\usepackage{color}
\begin{document}

\pagestyle{fancy}
\rhead{\includegraphics[width=2.5cm]{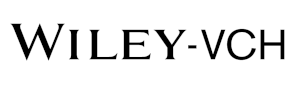}}

\title{Kinetics of ion migration in the electric field-driven manipulation of magnetic anisotropy of Pt/Co/oxide multilayers }

\maketitle

\author{Aymen Fassatoui,}
\author{Laurent Ranno,}
\author{Jose Pe\~{n}a Garcia,}
\author{Cristina Balan,}
\author{Jan Vogel,}
\author{H\'{e}l\`{e}ne B\'{e}a,}
\author{Stefania~Pizzini*}

\begin{affiliations}
Aymen Fassatoui, Laurent Ranno, Jose Pe\~{n}a Garcia, Cristina Balan, Jan Vogel, Stefania Pizzini \\
Univ.~Grenoble Alpes, CNRS, Institut N\'{e}el, Grenoble, France\\

H\'{e}l\`{e}ne B\'{e}a\\
Univ.~Grenoble Alpes, CEA, CNRS, Grenoble INP, IRIG-SPINTEC, Grenoble, France\\

* stefania.pizzini@neel.cnrs.fr
\end{affiliations}


\begin{abstract}
   Magneto-ionics, by which the magnetic properties of a thin layer can be modified through the migration of ions within a liquid or solid electrolyte,  is a fast developing research field. This is mainly due to the perspective of energy efficient magnetic devices, in which the magnetization direction is controlled not by a magnetic field or an electrical current, as done in traditional devices, but by an electric field, leading to a considerable reduction of energy consumption.    

  In this work, the interfacial perpendicular magnetic anisotropy (PMA)  of a series of Pt/Co/oxide trilayers covered by a ZrO$_2$ layer, acting as a ionic conductor, was finely tuned by a  gate voltage at room temperature. The non-volatility and the time evolution of the effect point at oxygen ion migration across the ZrO$_2$  layer as the driving mechanism. A large variation of the PMA is obtained by modifying the degree of oxidation of the cobalt layer with the flux of oxygen ions:  the easy magnetization axis  can be switched reversibly from in-plane, with an under-oxidized Co,  to in-plane, with an over-oxidized Co, passing through an out-of-plane magnetization with maximum PMA. The switching time between the different magnetic states is limited by the oxygen ion drift velocity through the multilayer structure. This was shown to depend exponentially on the applied bias voltage, and could be varied by over 5 orders of magnitude, from several minutes to a few ms.  On the other hand, for a fixed gate-voltage, the oxidation of the cobalt layer decreases  exponentially as a function of time.  This behavior is in agreement with the theoretical model developed  by Cabrera and Mott (1949) to explain the growth of  very thin oxides at low temperatures. The possibility to explain the observed effect with a relatively simple theoretical model opens the possibility to engineers materials with optimized properties.

\end{abstract}

\section{Introduction}
Voltage control of magnetic properties, and in particular magneto-ionics, is receiving a growing interest as it opens the possibility to largely improve the energetic efficiency of magnetic memories and other spintronic devices \cite{Nichterwitz2021}. Perpendicular magnetic anisotropy (PMA) is a key property of ultrathin magnetic films, as its strength determines the preferred orientation of their magnetization. The possibility to manipulate the PMA is therefore a key requirement for the conception of such devices. 

PMA in thin ferromagnetic (FM)  films is generally observed at interfaces with a heavy metal with large spin-orbit interaction, like Pt/Co, but surprisingly also at  FM-oxide interfaces \cite{Monso2002}. This is attributed to the hybridization between the metal and the oxygen electronic orbitals across the interfaces, as confirmed by ab initio calculations \cite{Yang2011}. The amplitude  of this interaction is strongly sensitive to the degree of oxidation of the  metal/oxide interface. In Pt/FM/MOx stacks, the maximum PMA is obtained for the optimal oxidation of the FM/oxide interface, when all the metal interfacial FM atoms being bonded to oxygen atoms. If the contribution of the Pt/FM interface to the PMA is not sufficient to promote out-of-plane magnetization, a slight increase or decrease of the oxidation (towards an over- or under-oxidized state)  may be sufficient to switch the magnetization from out-plane (OOP) to in-plane (IP) (see Figure \ref{fig:kinetics-Tb}(b)) \cite{Manchon2008a,Manchon2008b,Manchon2008c,Dieny2017}. 

The interfacial anisotropy at the FM-oxide interface can be manipulated by an electric field  via electronic \cite{Weisheit2007,Maruyama2009,Matsukura2015} or magneto-ionic effects \cite{Nichterwitz2021}. The largest effects of the electric field on the PMA were obtained by triggering the oxidation of cobalt via the migration of oxygen ions \cite{Bi2014,Bauer2015,Zhou2016}. Up to recent years, this mechanism was observed to be slow and to require high temperatures, since the voltage-driven switching between IP and OOP magnetization was shown to require several minutes at room temperature  \cite{Bi2014}. A voltage-driven  mechanism in which the modification of interfacial PMA is attributed to the migration of hydrogen, rather than to the heavier oxygen ions,  was  recently proposed as a more efficient \cite{Tan2019a,Tan2019b} and faster (down to 2 ms) \cite{Lee-Beach2020}  way to switch the magnetization. 

In this work, we show that by depositing on top of the magnetic stack a thin ZrO$_2$ layer acting  as an oxygen ion conductor, a huge voltage-driven variation of the PMA can be  obtained for two model systems  for spintronic applications,  Pt/Co/AlOx and Pt/Co/TbOx trilayers. While Pt/Co/AlOx \cite{Miron2011} was the first system in which chiral domain walls could be driven efficiently by spin-orbit torque, Pt/Co/TbOx may be interesting as    The mechanism, driven by oxygen ion migration that tunes the oxidation of the Co layer,  is non-volatile, reversible, and relatively robust. The switching time between OOP and IP magnetic state  depends exponentially on the applied bias voltage, and in Pt/Co/AlOx it  was decreased by over 5 orders of magnitude (from several minutes to a few ms) by increasing the bias voltages from 1 V and 6 V. 

The temporal variation of the interfacial anisotropy in the presence of an applied gate voltage can be modeled by assuming a parabolic variation of the PMA versus the oxidation rate, and an exponential variation of the oxidation/reduction of the cobalt layer versus time. The temporal variation of the oxidation rate is in agreement with the theory proposed by Cabrera and Mott in 1949 \cite{Cabrera1949} for the formation of a thin oxide film, which occurs  under the effect of the large electric field due to the contact potential difference at the oxide/metal interface.

\section{Experimental details}

\begin{figure*}[ht]
\begin{center}
\includegraphics[width=16cm]{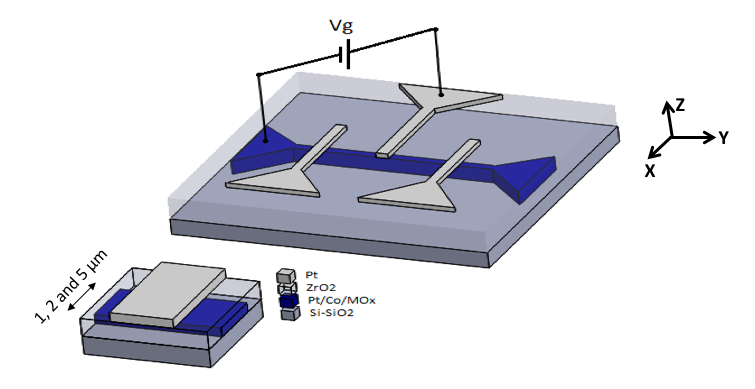}
\end{center}
\caption{\label{fig:device} \textbf{Schematic view of the device}. The Pt/Co/MOx layer is sputter deposited on a Si/SiO$_2$ substrate then patterned by EBL. A thin ZrO$_2$ layer is then deposited on the whole sample. Platinum electrodes are finally patterned, acting as local gates. }
\end{figure*}

Pt(4)/Co(1.1 to 1.2)/TbOx(0.6) and Pt(4)/Co(0.6)/AlOx(1) magnetic stacks were deposited by magnetron sputtering on Si/SiO$_2$ wafers (see also \cite{Fassatoui2020}). After patterning the films into 1-$\mu$m, 2-$\mu$m and 5-$\mu$m wide strips by electron beam lithography (EBL) and ion-beam etching,  a 10 nm thick ZrO$_2$ dielectric layer was deposited by atomic layer deposition (ALD). The oxide layer, grown at 100\textdegree C, has an amorphous structure. Finally, 6 nm thick Pt electrodes were patterned by EBL  and lift-off, acting as local gates (Figure \ref{fig:device}). 
To study the time evolution of the PMA during the application of the gate voltage, hysteresis loops were recorded using a polar magneto-optical Kerr magnetometer, with  the light focused on the gate electrode. The magnetic configuration of some of the magnetic stacks was measured by magnetic force microscopy  (MFM). 

\section{Non volatile manipulation of the PMA in Pt/Co/TbOx stacks}
\begin{figure*}[ht]
\begin{center}
\includegraphics[width=16cm]{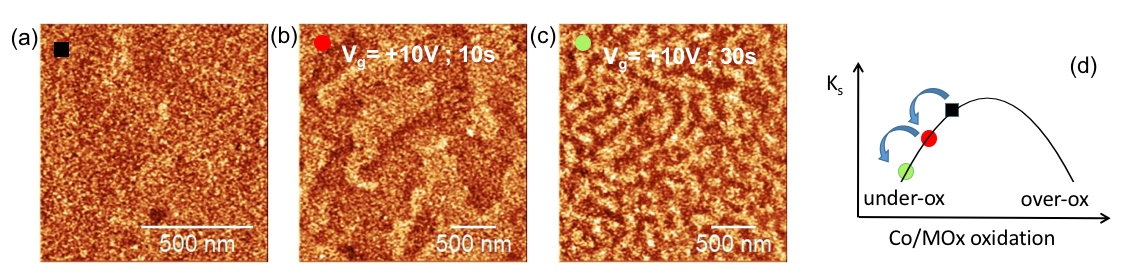}
\end{center}
\caption{\label{fig:MFM-Tb} \textbf{MFM images of the Pt/Co(1.2)/TbOx stack}.  MFM images showing the magnetic configuration of the Co layer in the initial state, where the magnetization is saturated OOP and the cobalt layer is under-oxidized (less oxidized than for  the maximum PMA)  (a) and after the application  of a bias voltage V$_{g}$=+10~V for 10~s (b) and 30~s (c). The formation of labyrinthine domains of increasing size indicates the progressive decrease of the domain wall energy due to the  decrease of the PMA. The images are scanned from left to right, and from bottom to top. }
\end{figure*}

The magnetization configuration of Pt/Co(1.2)/TbOx was observed by MFM both in the as-patterned state, characterised by a saturated out-of-plane magnetization with an  under-oxidized Co layer, and after the application of a constant bias voltage of +10~V for 10~s and 30~s, at room temperature. Each image was taken with a closed circuit,  after the removal of the bias voltage. The three images are shown in Figure \ref{fig:MFM-Tb}(a-c).

In the initial state, the MFM image (Figure \ref{fig:MFM-Tb}(a)) shows as expected  a homogeneous contrast.
 When the bias field is applied for 10~s (Figure \ref{fig:MFM-Tb}(b)), labyrinthine magnetic domains appear, with an average width of the order of 400~nm. 
 When the same positive voltage is applied for 20 more seconds (Figure \ref{fig:MFM-Tb}(c))  the width of the magnetic domains decreases to around 100~nm.  Since the formation of a demagnetized state reveals that the systems  is close to the re-orientation transition between OOP and IP magnetization, our results show that the positive voltage leads to a decrease of the PMA.  


 The labyrinthine domain width $L$ in the case of ultrathin ferromagnetic layers is given by \cite{Schafer1998,Kaplan1993} $L=C~t~\exp(\pi L_{0}/t)$, where $L_{0}=\sigma/\mu_{0}M_{s}^{2}$ is the characteristic dipolar length, $\sigma$ is the domain wall energy, $t$ is the ferromagnetic film thickness and $C$ is a numerical constant of the order of 1. In the non-centrosymmetric stacks studied in this work, domain walls have chiral N\'{e}el structure \cite{Boulle2016,Juge2019} and their energy is given by $\sigma=4\sqrt{AK_{eff}}-\pi |D|$,  where $A$ is the exchange stiffness and  $D$ is the strength of the Dzyaloshinskii-Moriya interaction. The domain width then depends exponentially on the domain wall energy, and therefore on the effective magnetic anisotropy. The observed decrease of the domain size for increasing duration of the applied bias reflects the progressive decrease  of the interfacial PMA. 
 
The domain width observed in Figs. \ref{fig:MFM-Tb} (b,c) does not evolve during the  MFM scan time (8 minutes) and for several days after the application of the voltage: we then define the magnetic state obtained after the application of the bias voltage as non-volatile. 
As already illustrated by our previous work \cite{Fassatoui2020}, the cumulative and non-volatile variation of the magnetic anisotropy leads us to exclude charge accumulation/depletion effects and to suggest that the voltage-driven migration of oxygen ions is the most probable driving mechanism. Since a positive bias drives oxygen ions away from the  Co layer, this leads to a decrease of its oxidation which causes a decrease of the PMA  (see sketch in Figure \ref{fig:MFM-Tb}(d)).

\section{Large tuning of interfacial anisotropy}

\begin{figure*}[ht]
\begin{center}
\includegraphics[width=16cm]{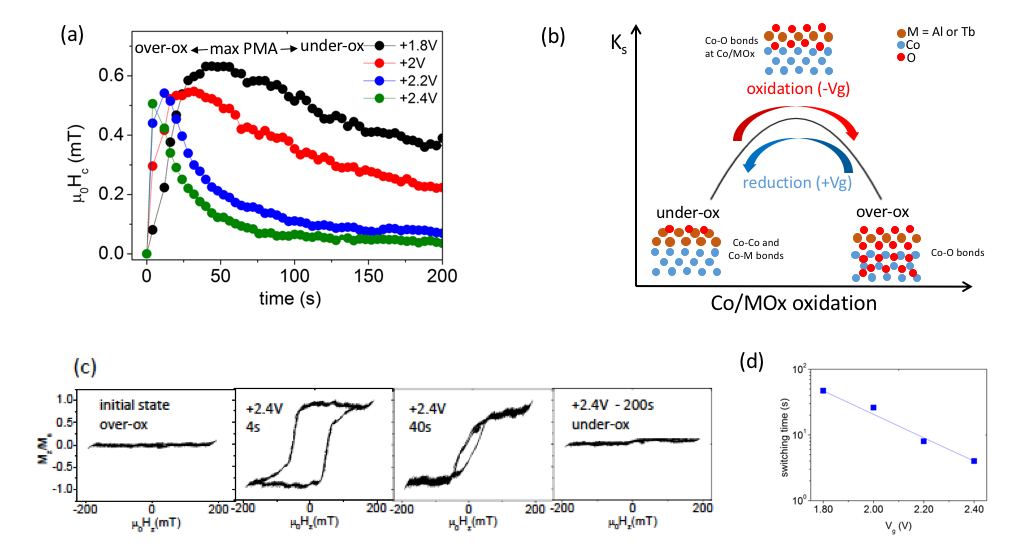}
\end{center}
\caption{\label{fig:kinetics-Tb} \textbf{Large variation of PMA in Pt/Co(1.1)/TbOx stack under positive bias}. (a) The coercive field measured by polar MOKE with the light focused on the top Pt electrode is plotted versus the time of application $t$ of the bias voltage V$_{g}$.  Starting from an over-oxidized cobalt layer with in-plane magnetization, the positive bias voltage drives the reduction of  the Co layer, leading to a change of the PMA; (b) sketch of mechanism leading to the variation of the PMA through the oxidation/reduction of the Co layer; (c) some examples of hysteresis loops recorded in the over-oxidized initial state with IP magnetization, and at several stages (t=4s, 40s and 200s) during the application of V$_g$=+2.4V; (d) switching time $t$ from IP (over-oxidized initial state) to OOP (maximum PMA state)  as a function of bias voltage V$_{g}$.  }
\end{figure*}

The results just described show that the magnetic anisotropy can be finely tuned by adjusting the application time of the bias voltage and therefore the oxidation of the cobalt layer. 
Based on the results of previous investigations on Pt/Co/oxide systems \cite{Monso2002,Manchon2008a,Manchon2008b,Manchon2008c,Dieny2017}  we anticipate that the voltage-driven migration of oxygen ions might be used to tune the magnetization  from IP (for an under-oxidized Co)  
 to IP (for an over-oxidized Co) passing through a state with maximum PMA (see sketch in Figure \ref{fig:kinetics-Tb}(b)). 
 
  This was partially achieved by Bi \textit{et al.} \cite{Bi2014} for Pt/Co/Gd$_{2}$O$_{3}$ stacks where the electric field was applied at  high temperature  (260\textdegree C) and for long times (up to 600~s).   
Our results are shown in Figure \ref{fig:kinetics-Tb}(a) for a Pt/Co(1.1)/TbOx 5x5 $\mu m^{2}$ capacitor-like structure. (Note that in this stack the Co layer is slightly thinner than in the one described in the previous section).  Starting from an over-oxidized state with IP 
 magnetization, the 
time evolution of the coercive field  $\mu H_C$ was measured during the application of
 several positive bias voltages V$_{g}$ (from +1.8~V to +2.4~V) which reduces the Co layer.
Each point in the curve represents the average coercivity measured over 4 hysteresis cycles, each taking 1 second.  
In these curves, the variation of $\mu H_C$ reflects the variation of the anisotropy at the Co/oxide interface \cite{Givord2003}, so we will discuss the results in terms of the PMA. For each bias voltage, the PMA  first increases, so that the  magnetization switches from IP to OOP,  passes through a maximum (optimum oxidation) and eventually decreases as the Co interface becomes under-oxidized and the magnetization turns in-plane (as shown by the hysteresis cycles in Figure \ref{fig:kinetics-Tb}(c) and sketched in Figure \ref{fig:kinetics-Tb}(b)). A large change of the interfacial anisotropy (estimated $\beta >$ 6000 fJ/(Vm))  is therefore obtained using reversible oxygen ion migration at room temperature as the driving mechanism. 
 
 For each bias voltage V$_{g}$, the transition between the initial IP over-oxidized state and
 the maximum PMA state is faster than the transition 
 between the maximum PMA state and under-oxidized IP state.  The two transition times decrease exponentially with the bias voltage amplitude, 
 as shown in Figure \ref{fig:kinetics-Tb}(d) for the IP (over-oxidized state) to the OOP state with maximum PMA. By increasing the bias voltage from +1.8~V 
 to +2.4~V, the switching time at room
  temperature decreases from 48~s to 4~s. As will be discussed later, this behaviour 
 is in agreement with an exponential
  increase of the drift velocity of oxygen ions versus V$_{g}$  expected for the very large electric fields used in this work \cite{Strukov2009}. 
  

\section{Stability of the under-oxidized magnetic state and spontaneous re-oxidation }

When the  positive  bias voltage is removed after reaching an under-oxidized state with IP magnetization, our measurements on the Pt/Co(1.1)/TbOx stack suggest the occurring of a spontaneously re-oxidation of the Co top interface.  This is shown in Figure \ref{fig:relaxation-Tb} where we present the temporal evolution of the coercive field after the removal of the bias voltage  V$_{g}$=+2.4V, applied for t=200~s. Starting from the IP magnetization (under-oxidized state induced by the application of the positive gate voltage), the coercivity (and therefore the PMA)
is observed to quickly increase, to reach a maximum and  then to decrease and vanish.  
The final equilibrium state of the system 
is characterised by an over-oxidized Co layer with IP magnetization.

This behavior is different from the one obtained for the Pt/Co(1.2)/TbOx stack described above (Figure \ref{fig:MFM-Tb}),  where the stripe domain structure in the under-oxidized state was observed to be stable over a long time. 
The two Pt/Co/TbOx stacks, grown simultaneously with a Co thickness gradient, differ only for the slight variation of the Co thickness. 
The labyrinthine structure characterising the demagnetized state is obtained when $K_{eff}= (K_{s}^{Pt} +K_{s}^{ox})/t_{Co}-1/2\mu_{0}M_{s}^{2}\simeq 0$, where 
 $K_{s}^{Pt}$ and $K_{s}^{ox}$ are respectively the interfacial magnetic anisotropies at the Pt/Co and Co/oxide interfaces. The critical $K_{s}^{*ox}$ satisfying this condition decreases as $t_{Co}$ decreases. Therefore, a  stripe domain configuration with similar domain size is obtained for a more oxidized Co interface for $t_{Co}$=1.2~nm than for $t_{Co}$=1.1~nm. 

The difference in the stability of the two labyrinthine domain structures may therefore be explained by the presence of a critical Co oxidation state $x^*$ below which the magnetic state becomes unstable and the Co spontaneously re-oxidizes: for $t_{Co}$=1.2~nm the labyrinthine domain state is obtained for  $x > x^*$, leading to a stable magnetic texture, while for $t_{Co}$=1.1~nm $x < x^*$ which leads to an unstable magnetic configuration. 
The underlying microscopic mechanism will be discussed later.

\section{Modeling the time variation of the interfacial PMA }

\begin{figure*}[ht]
\begin{center}
\includegraphics
[width=16cm]{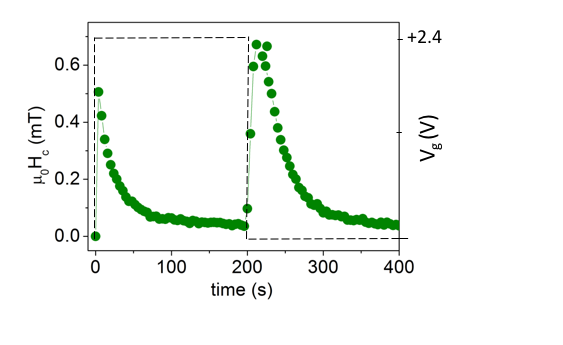}
\end{center}
\caption{\label{fig:relaxation-Tb} \textbf{Spontaneous re-oxidation of the Co layer in the Pt/Co(1.1)/TbOx stack}.  Coercive field versus time showing the re-oxidation of the Co layer after the removal of  the positive bias V$_{g}$=+2.4~V, as in Figure \ref{fig:kinetics-Tb}(a), after $t$=200 s.}


\newpage

\end{figure*}

We can note that the time evolution of the coercive field during 
the spontaneous re-oxidation of the Co layer (Figure \ref{fig:relaxation-Tb}) is very similar to that observed when the Co interface is reduced under a positive voltage (Figure \ref{fig:kinetics-Tb}(a)). 
This indicates that a common mechanism might govern the electric field driven reduction and the spontaneous re-oxidation of the Co interface.
In order to 
explain the temporal variation of the PMA during the application of a constant bias voltage, some 
assumption has to be made on the dependence of the PMA on the oxidation of the Co layer, and on the time evolution of the Co oxidation.

\begin{figure*}[ht]
\begin{center}
\includegraphics[width=16cm]{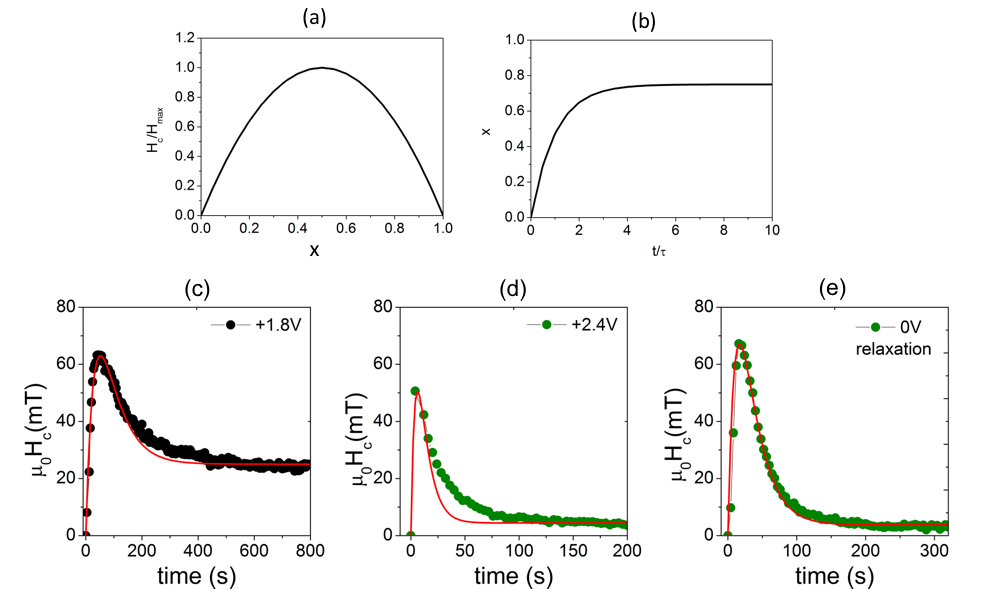}
\end{center}
\caption{\label{fig:model} \textbf{Modeling the temporal evolution of the PMA of Pt/Co(1.1)/TbOx during the application of a bias voltage.} (a) proposed parabolic variation of the PMA as a function of the oxidation rate $x$ of the cobalt layer; (b) proposed time evolution of the oxidation rate of the Co layer, decreasing exponentially versus time; (c-d) coercivity versus time measured for the Pt/Co(1.1)/TbOx stack for V$_{g}$=+1.8~V (c) and V$_{g}$=+2.4~V (d) and the corresponding fits using Equation 3; (e) coercive field versus time showing the re-oxidation of the Co layer after the removal of  the positive bias V$_{g}$=+2.4~V, and the corresponding fit with Equation 3.    }
\end{figure*}

Previous studies carried out for several Pt/Co/oxide stacks \cite{Monso2002,Manchon2008a,Manchon2008b,Manchon2008c,Dieny2017} show that the PMA changes with the cobalt oxidation following a  typical bell-like curve, with a maximum corresponding to the optimum oxidation state. This curve may be approximated by  a parabola, as sketched in Figure \ref{fig:model}(a), where the oxidation rate is represented by the parameter $x$:  here $x$=0 corresponds to the IP-OOP
transition; $x$=1/2 corresponds to optimum oxidation and maximum PMA; $x$=1, corresponds to the OOP-IP transition. We assume that the model is valid both for the oxidation and the reduction mechanism.
We also assume that the coercive field is proportional to the anisotropy K$_{eff}$ \cite{Givord2003} and that this
has a quadratic dependence on $x$ i.e. that K$_{eff}$(x) = 4 K$_{max}x(1-x)$. We then have (Figure \ref{fig:model}(a)):
\begin{equation}
\frac{H_{_c}(x)}{H_{max}}= 4x(1-x)
\end{equation}

The applied electric field modifies the oxidation of the cobalt layer and we assume that the oxidation rate follows an exponential law with 
a characteristic time $\tau$ (Figure \ref{fig:model}(b)). The system is always prepared (reset) with
$x$=0 at $t$=0 (IP-OOP transition). The oxidation state observed at time $t$  depends on the strength of the electric field and will be called $x_{final}$.
\begin{equation}
x(t)= x_{final}(1-e^{-t/\tau})
\end{equation}
By putting H$_{c}(x)$ and $x(t)$ together we obtain:
 \begin{equation}
\frac{H_{c}}{H_{max}}=4x_{final}(1-e^{-t/\tau})(1-x_{final}(1-e^{-t/\tau}))
\end{equation}
where the maximum of $H_{c}$ is obtained for $t \approx \tau$.

In Figures \ref{fig:model}(c-d) the experimental variations of the coercivity versus time obtained for bias voltage V$_{g}$=+1.8~V and +2.4~V  
(Figure \ref{fig:kinetics-Tb}(a)) are fitted with the expression in Equation 3, using $x_{final}$ and $\tau$ as fitting parameters.
The proposed model appears to describe well the experimental curve obtained with the lower bias voltage (V$_{g}$=+1.8~V), except for a small deviation observed for coercive fields smaller than 40 mT in the under-oxidized part of the curve. 
For V$_{g}$=+2.4~V the model fails to describe the part of the curve with 0 $\leq$ H$_{c}$ $\leq$ 40 mT in the  under-oxidized part of the curve, where the decay of the coercivity is much lower than that described by Equation 3.
Following the discussion of the  previous section, this behaviour might be explained by the spontaneous re-oxidation of the cobalt layer occurring, for $x < x^*$, through the diffusion of oxygen towards the 
Co interface, therefore counteracting the effect of the positive bias voltage.

Interestingly, Figure \ref{fig:model}(e) shows that the experimental coercivity versus time curve describing the re-oxidation of the Co interface after removal 
of the bias voltage (as in Figure \ref{fig:relaxation-Tb}(c)) can be perfectly fitted with the oxidation rate law described by our model. This results is in agreement 
with the theory developed in 1949 by Cabrera and Mott who proposed that the growth of very thin oxide layers at room temperature occurs following an exponential law, similarly to Equation 2 \cite{Cabrera1949}. As far as we know, the "real time" variation of PMA of a magnetic stack occurring during the oxidation of a metallic layer  has not been observed before.

\section{Exponential decrease of  the switching time between IP and saturated OOP states}

The  voltage-driven oxidation/reduction kinetics was also studied for a Pt/Co(0.6)/AlOx sample. In this case the measurements were carried out on a 2x2 $\mu m^{2}$ capacitor-like structure. In the initial state (time $t$=0) the Co layer is over-oxidized with vanishing coercivity and IP magnetization. 

\begin{figure*}[ht]
\begin{center}
\includegraphics[width=16cm]{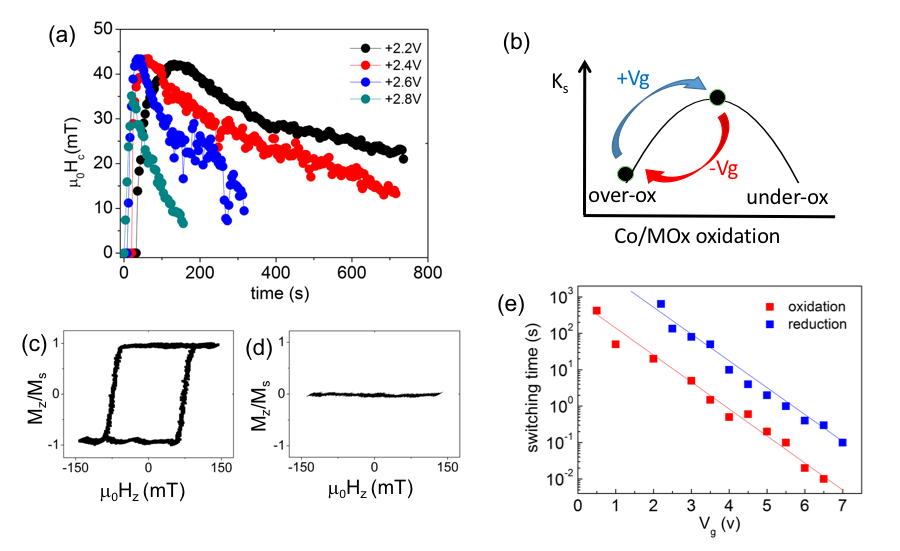}
\end{center}
\caption{\label{fig:PtCoAlOx} \textbf{Large variation of the PMA in Pt/Co(0.6)/AlOx stack under positive bias and electric field dependence of the IP-OOP switching time} (a): The coercive field measured by polar MOKE with the light focused on the top Pt electrode is plotted versus the time of application $t$ of the bias voltage V$_{g}$.  Starting from an over-oxidized cobalt layer with in-plane magnetization, the positive bias voltage drives the reduction of  the Co layer, leading to a change of the PMA; (b) sketch of the two magnetic states between which the PMA has been switched for different values of the positive and negative gate voltage; the state with maximum PMA and a state with IP magnetization. The same states can be easily re-initialised with the gate voltage, by recording the corresponding hysteresis loops; (c-d) hysteresis loops measured in the two magnetic states, sketched in (b), between which the PMA has been switched; (d) duration of the gate voltage pulse $t$  needed to switch between IP over-oxidized magnetization and OOP magnetization with maximum PMA (with negative voltage) and viceversa (with positive gate voltage) as a function of bias voltage V$_{g}$. Note that the gate voltages are negative for the red curve (leading to oxidation of the Co layer) and positive for the blue curve (leading to reduction of the Co layer).  }
\end{figure*}

The curves in Figure \ref{fig:PtCoAlOx}(a) show the variation of the coercive field as a function of time, during the application of several positive bias voltages driving oxygen ions away from the Co/oxide interface. Similarly to Pt/Co/TbOx, for every bias voltage the coercivity, and therefore the interfacial PMA,  increases rapidly up to the maximum, then decreases slowly in the part of the curve where Co is under-oxidized. 

The duration of the gate voltage pulse needed to switch the magnetization between the initial over-oxidized state with IP magnetization  and the state with maximum PMA state and viceversa (Figure \ref{fig:PtCoAlOx}(b-c)) were measured, for positive and negative bias respectively (Figure \ref{fig:PtCoAlOx}(d)).  The switching times are observed to change by 5 orders of magnitude in the used voltage range, e.g. in the oxidation process, the switching time varies between 420 s for -0.5V to 10 ms for -6.5 V. 
In both the oxidizing and the reducing process and over the whole time range, the switching time $\tau$ decreases exponentially with the bias voltage. Moreover, with an equivalent (and opposite) bias voltage, the oxidation process is faster than the reduction.  

\section{Discussion}
Let's summarise the main results of this work:

I) the switching time $\tau$  between an over-oxidized state with in-plane anisotropy (corresponding to the formation of an Co oxide layer of a certain thickness $t_{ox}$) and the state with OOP magnetization with  maximum PMA (which we associate to a state with Co covered with one monolayer of oxygen atoms) decreases exponentially with the applied gate voltage. 

II) the time dependence of the PMA (i.e. of the oxidation/reduction of the Co layer) can be well described by Equation 2, that supposes an exponential decrease of the oxidation/reduction rate as a function of time.  This is the case both for the bias-driven oxidation/reduction and for the spontaneous re-oxidation of the under-oxidized Co layer in Pt/Co(1.1)/TbOx, once the bias is removed.


Both the exponential variation of the switching time  $\tau$ and the exponential time-dependence of the oxidation rate of the cobalt layer can be  explained by the atomistic model developed  by  Cabrera and Mott in the 1940's \cite{Cabrera1949} to explain the growth rate of very thin oxide films. According to the authors, the rate of oxidation is limited by the injection of defects (i.e. the migration of ionic species) within the oxide layer and across the metal/ion interface. These ionic species are driven by the electric field $E$ present at the interface between a metal and the thin oxide, due to the contact potential difference $V$ between the metal and the adsorbed oxygen ion layer. According to the authors, for thin oxide layers of thickness $t_{ox} \ll 1\mu m$, the electric field $E=V/t_{ox}$, can be so strong that the drift velocity of the ions is no longer linear with $E$, but changes exponentially with it.    

From Refs. \cite{Cabrera1949, Strukov2009},  for strong electric fields  ($E \gg E_{0}$), the ion drift velocity reads: 
\begin{equation}
    v_{ion} = 2fa e^\frac{-U_A}{k_{B}T }  e^\frac{E}{E_{0} }
    \label{Eq:ion-speed}
\end{equation}
   where $E_{0}  = 2 k_{B}T / (qa)$ is the characteristic field for the ion (typically
1 MV/cm for T =  300 K), $q$ is the electrical charge of the ion, $f$ is the attempt frequency for the ion jumps (typically 10$^{12}$-10$^{13}$ Hz), $a$ is the periodicity of the ion vacancy sites in the material and $ U_A$ is the activation energy of the ion jump from one site to the next. 

From Equation \ref{Eq:ion-speed} follows that the rate of growth of the oxide film decreases exponentially with the film thickness, as the electric field $E$ is  proportional to 1/$t_{ox}$. 

These arguments may be applied  to our system. Let us first consider the results shown in Figure \ref{fig:PtCoAlOx}.  If we assume that the switching time $\tau$  is mainly limited by the drift velocity of the oxygen ions across the ZrO$_2$ layer ($t_{ZrO2}$=10 nm), so that  $v_{ion} = t_{ZrO2}/\tau$, the exponential variation of $v_{ion}$ obtained with fields of the order of 1 MV/cm is in agreement with the theorical model just described.  
From the slope of the experimental velocity curves for negative and positive bias (oxidation and reduction processes, Figure \ref{fig:PtCoAlOx}), we obtain $E_0  \approx$   0.6 x 10$^{6}$ V/cm,  which is in the range expected for the non-linear ionic transport. The best fit of the two curves to Equation \ref{Eq:ion-speed} gives $U_A \approx $1 eV, with a small difference between the activation energies for negative and positive bias,  $U_{A,oxidation}-U_{A,reduction}$ $\approx$ 3.3 $k_{B}T$, that can also be obtained from the shift of the two curves.  

A question arises on whether the rate-limiting barrier is situated in the dielectric layer or at the metal/oxide interface. The small difference (with respect to the enthalpy of formation of Co oxide)  between the barriers for positive and negative bias voltages, leads us to propose that the switching time is limited by the barrier associated to ionic migration in ZrO$_2$. The value of $U_A \approx$1 eV obtained in this work is of the same order of magnitude of that recently  calculated for oxygen ion migration via oxygen vacancies in monoclinic ZrO$_2$ \cite{Hur2020}.  The lower activation barrier, compared to experimental values found in the literature ($\approx$ 2 eV) may be associated to the amorphous sub-stoichiometric nature of our oxide, which was grown by ALD at low temperature to avoid deteriorating the magnetic properties of the magnetic stacks. 


The curve describing the natural re-oxidation of the Co layer after the removal of the bias voltage (Figs. \ref{fig:relaxation-Tb} and \ref{fig:model}(e), that we fitted assuming an exponential decrease of the oxidation rate as a function of time, is also  in agreement with the model developed by Cabrera and Mott, that predicts an exponential variation of the growth rate of the oxide as a function of its thickness. It is interesting to note that the characteristic time for the re-oxidation of the Co layer (with V$_g$=0)  is comparable to that obtained for the Co oxidation under a bias voltage V$_g$=+1.5V (E=1.5 MV/cm). This seems to prove that the electric field at the metal/oxide interface is indeed extremely large, of the order of some 1 MV/cm. 


\section{Conclusion}

We have shown that a ZrO$_2$ layer deposited on top of  Pt/Co/MOx microstructures acts as a ionic conductor that, under the action of a gate voltage, draws oxygen ions towards/away from the Co layer, modifying its oxidation and therefore the anisotropy of the magnetic stacks. The effect on the magnetic properties is large and reversible. The temporal stability of the effect depends on the oxidation state of the Co layer after the application of the gate voltage i.e. we have observed that strongly under-oxidized Co layers in Pt/Co(1.1)/TbOx stacks spontaneously re-oxidize. The temporal variation of the anisotropy during the application of the gate voltage has been explained supposing a parabolic variation of the PMA and an exponential decrease of the oxidation/reduction as a function of time. 
We have shown that the voltage-driven switching time between different magnetic configurations decreases exponentially with the applied voltage, and can be decreased by several orders of magnitude, down to a few ms, by tuning the amplitude of the electric field between $\approx$1 and $\approx$6 MV/cm. The microscopic mechanism can be explained by a theoretical model predicting the exponential variation of the ion drift velocity for very large electric fields. The possibility to explain the observed effect with a relatively simple theoretical model opens the possibility to engineers materials with optimized properties.

\section{Acknowledgements}

We acknowledge the support of the Agence Nationale de la Recherche (projects ANR-17-CE24-0025 (TOPSKY), ANR-16-CE24-0018 (ELECSPIN) and ANR-19-CE24-0019 ADMIS)) and of the DARPA TEE program through Grant No. MIPR HR0011831554. The authors acknowledge funding from the European Union’s Horizon 2020 research and innovation program under Marie Sklodowska-Curie Grant Agreement No. 754303 and No. 860060 “Magnetism and the effect of Electric Field” (MagnEFi). J.P.G. also thanks the Laboratoire d\textquotesingle Excellence LANEF in Grenoble (ANR-10-LABX-0051) for its support. B. Fernandez, T. Crozes, Ph. David, E. Mossang and E. Wagner are acknowledged for their technical help.  A.F. thanks O. Fruchart for his support for the MFM experiments. S.P. thanks M. Maroun and Ph. Allongue for the many inspiring discussions. 

\bibliographystyle{unsrt}

\begin{thebibliography}{10}

\bibitem{Nichterwitz2021}
M.~Nichterwitz, S.~Honnali, M.~Katuzau, J.~Zehner, K.~Nielsch, and K.~Leistner.
\newblock Advances in magneto-ionic materials and perspectives for their
  application.
\newblock {\em APL Mater.}, 9:030903, 2021.

\bibitem{Monso2002}
S.~Monso, B.~Rodmacq, S.~Auffret, G.~Casali, F.~Fettar, B.~Gilles, B.~Dieny,
  and P.~Boyer.
\newblock Crossover from in-plane to perpendicular anisotropy in {
  Pt/CoFe/AlOx} sandwiches as a function of { Al } oxidation: A very accurate
  control of the oxidation of tunnel barriers.
\newblock {\em Appl. Phys. Lett.}, 80:4157, 2002.

\bibitem{Yang2011}
H.~X. Yang, M.~Chshiev, B.~Dieny, J.~H. Lee, A.~Manchon, and K.~H. Shin.
\newblock First-principles investigation of the very large perpendicular
  magnetic anisotropy at { Fe/MgO and Co/MgO} interfaces.
\newblock {\em Phys. Rev. B}, 84:054401, 2011.

\bibitem{Manchon2008a}
A.~Manchon, C.~Ducruet, L.~Lombard, S.~Auffret, B.~Rodmacq, B.~Dieny,
  S.~Pizzini, J.~Vogel, V.~Uhl\'{\i}\u{r}, M.~Hochstrasser, and G.~Panaccione.
\newblock Analysis of oxygen induced anisotropy crossover in {Pt/Co/MO$_{x}$}
  trilayers.
\newblock {\em J. Appl. Phys.}, 104:043914, 2008.

\bibitem{Manchon2008b}
A.~Manchon, S.~Pizzini, , J.~Vogel, V.~Uhl\'{\i}\u{r}, L.~Lombard, C.~Ducruet,
  S.~Auffret, B.~Rodmacq, B.~Dieny, M.~Hochstrasser, and G.~Panaccione.
\newblock X-ray analysis of oxygen-induced perpendicular magnetic anisotropy in
  {Pt/Co/AlO$_{x}$} trilayers.
\newblock {\em J. Magn. Magn. Mater.}, 320:1889, 2008.

\bibitem{Manchon2008c}
A.~Manchon, S.~Pizzini, V.~Vogel, J.~Uhl\'{\i}\u{r}, L.~Lombard, C.~Ducruet,
  S.~Auffret, B.~Rodmacq, B.~Dieny, M.~Hochstrasser, and G.~Panaccione.
\newblock X-ray analysis of the magnetic influence of oxygen in
  {Pt/Co/AlO$_{x}$} trilayers.
\newblock {\em J. Appl. Phys.}, 103:07A912, 2008.

\bibitem{Dieny2017}
B.~Dieny and M.~Chshiev.
\newblock Perpendicular magnetic anisotropy at transition metal/oxide
  interfaces and applications.
\newblock {\em Rev. Mod. Phys.}, 89:025008, 2017.

\bibitem{Weisheit2007}
M.~Weisheit, S.~F{\"a}hler, A.~Marty, Y.~Souche, C.~Poinsignon, and D.~Givord.
\newblock Electric field-induced modification of magnetism in thin-film
  ferromagnets.
\newblock {\em Science}, 315:349, 2007.

\bibitem{Maruyama2009}
T.~Maruyama, Y.~Shiota, T.~Nozaki, K.~Ohta, N.~Toda, M.~Mizuguchi, A.~A.
  Tulapurkar, T.~Shinjo, M.~Shiraishi, and S.~Mizukami.
\newblock Large voltage-induced magnetic anisotropy change in a few atomic
  layers of iron.
\newblock {\em Nat. Nanotech.}, 4:158, 2009.

\bibitem{Matsukura2015}
F.~Matsukura, Y.~Tokura, and H.~Ohno.
\newblock Control of magnetism by electric fields.
\newblock {\em Nat. Nanotech.}, 10:209, 2015.

\bibitem{Bi2014}
C.~Bi, Y.~Liu, T.~Newhouse-Illige, M.~Xu, M.~Rosales, J.~W. Freeland,
  O.~Mryasov, S.~Zhang, S.~G.~E. Velthuis, and W.~G. Wang.
\newblock Reversible control of {Co} magnetism by voltage induced oxidation.
\newblock {\em Phys. Rev. Lett.}, 113:267202, 2014.

\bibitem{Bauer2015}
U.~Bauer, Y.~Lide, J.~T. Aik, P.~Agrawal, S.~Emori, H.L. Tuller, S.~van Dijken,
  and G.S.D. Beach.
\newblock Magneto-ionic control of interfacial magnetism.
\newblock {\em Nat. Mater.}, 14:174, 2015.

\bibitem{Zhou2016}
X.~Zhou, Y.~Yan, M.~Jiang, B.~Cui, F.~Pan, and C.~Song.
\newblock Role of oxygen ion migration in the electrical control of magnetism
  in {Pt/Co/Ni/HfO$_{2}$} films.
\newblock {\em The Journal of Physical Chemistry C}, 120:1633, 2016.

\bibitem{Tan2019a}
A.J. Tan, M.~Huang, S.~Sheffels, F.~B\"{u}ttner, S.~Kim, A.H. Hunt, I.~Waluyo,
  H.L. Tuller, and G.~S.~D. Beach.
\newblock Hydration of gadolinium oxide ({GdO$_{x}$}) and its effect on
  voltage-induced {Co} oxidation in a {Pt/Co/GdO$_{x}$/Au} heterostructure.
\newblock {\em Phys. Rev. Materials}, 3:064408, 2019.

\bibitem{Tan2019b}
A.J. Tan, M.~Huang, C.O. Avci, F.~B\"{u}ttner, M.~Mann, W.~Hu, C.~Mazzoli,
  S.~Wilkins, H.L. Tuller, and G.~S.~D. Beach.
\newblock Magneto-ionic control of magnetism using a solid-state proton pump.
\newblock {\em Nat. Mater.}, 18:35, 2019.

\bibitem{Lee-Beach2020}
Ki-Young Lee, Sujin Jo, Aik~Jun Tan, Mantao Huang, Dongwon Choi, Jung~Hoon
  Park, Ho-Il Ji, Ji-Won Son, Joonyeon Chang, Geoffrey S.~D. Beach, and
  Seonghoon Woo.
\newblock Fast magneto-ionic switching of interface anisotropy using
  yttria-stabilized zirconia gate oxide.
\newblock {\em Nano Letters}, 20:3435, 2020.
\newblock PMID: 32343588.

\bibitem{Cabrera1949}
N.~Cabrera and N.F. Mott.
\newblock Theory of the oxidation of metals.
\newblock {\em Rep. Prog. Phys.}, 12:163, 1949.

\bibitem{Fassatoui2020}
Aymen Fassatoui, Jose Pe\~na Garcia, Laurent Ranno, Jan Vogel, Anne
  Bernand-Mantel, H\'el\`ene B\'ea, Sergio Pizzini, and Stefania Pizzini.
\newblock Reversible and irreversible voltage manipulation of interfacial
  magnetic anisotropy in $\mathrm{Pt}$/$\mathrm{Co}$/oxide multilayers.
\newblock {\em Phys. Rev. Applied}, 14:064041, 2020.

\bibitem{Schafer1998}
A.~Hubert and R.~Sch{\"a}fer.
\newblock {\em Magnetic Domains, The Analysis of Magnetic Microstructures}.
\newblock Springer: Berlin, 1998.

\bibitem{Kaplan1993}
B.~Kaplan and G.~A. Gehring.
\newblock The domain structure in ultrathin magnetic films.
\newblock {\em J. Magn. Magn. Mater.}, 128:111, 1993.

\bibitem{Boulle2016}
O.~Boulle, J.~Vogel, H.~Yang, S.~Pizzini, D.~de~Souza~Chaves, A.~Locatelli,
  T.O. Mentes, A.~Sala, L.~D. Buda-Prejbeanu, O.~Klein, M.~Belmeguenai,
  Y.~Roussign\'e, Y., A.~Stashkevich, S.M. Ch\'erif, L.~Aballe, M.~Foerster,
  M.~Chshiev, S.~Auffret, I.M. Miron, and G.~Gaudin.
\newblock {Room-temperature chiral magnetic skyrmions in ultrathin magnetic
  nanostructures}.
\newblock {\em Nat. Nanotech.}, {11}:{449}, {2016}.

\bibitem{Juge2019}
R.~Juge, S.-G. Je, D.~de~Souza~Chaves, L.~Buda-Prejbeanu, J.~Pe\~{n}a Garcia,
  J.~Nath, I.M. Miron, K.G. Rana, L.~Aballe, M.~Foerster, F.~Genuzio, F.O.
  Mentes, A.~Locatelli, F.~Maccherozzi, S.S Dhesi, M.~Belmeguenai,
  Y.~Roussign\'e, S.~Pizzini, G.~Gaudin, J.~Vogel, and O.~Boulle.
\newblock {Current-Driven Skyrmion Dynamics and Drive-Dependent Skyrmion Hall
  Effect in an Ultrathin Film}.
\newblock {\em Phys. Rev. Applied}, 12:044007, 2019.

\bibitem{Givord2003}
Dominique Givord, Michel Rossignol, and Vitoria~M.T.S. Barthem.
\newblock The physics of coercivity.
\newblock {\em Journal of Magnetism and Magnetic Materials}, 258-259:1--5,
  2003.
\newblock Second Moscow International Symposium on Magnetism.

\bibitem{Strukov2009}
D.B. Strukov and R.S. Williams.
\newblock Exponential ionic drift: fast switching and low volatility of
  thin-film memristors.
\newblock {\em Appl. Phys A}, 94:515, 2009.

\bibitem{Hur2020}
J.H. Hur.
\newblock First principles study of oxygen vacancy activation energy barrier in
  zirconia-based resistive memory.
\newblock {\em Sci. Rep.}, 10:5405, 2020.

\end{thebibliography}

\newpage

\newpage

\newpage
  
\newpage


\newpage

\end{document}